\documentclass[fleqn,usenatbib]{mnras}

\usepackage[T1]{fontenc}

\DeclareRobustCommand{\VAN}[3]{#2}
\let\VANthebibliography\thebibliography
\def\thebibliography{\DeclareRobustCommand{\VAN}[3]{##3}\VANthebibliography}

\usepackage{graphicx}	
\usepackage{amsmath,amssymb}	
\usepackage{multirow}
\usepackage{makecell}
\usepackage{orcidlink}
\usepackage{physics}
\usepackage{pdflscape}
\usepackage{xcolor}
\usepackage{newtxtext,newtxmath}

\title[
    \textit{Quasar Island} -- Three new $z\sim$ 6 quasars
]{
    \textit{Quasar Island} -- Three new $z\sim6$ quasars, including a lensed candidate, identified with contrastive learning
}

\author[X. Byrne et\ al.]{
Xander Byrne
\orcidlink{0000-0001-9488-238X}$^{1,2}$\thanks{E-mail: ajnb3@ast.cam.ac.uk},
Romain A. Meyer
\orcidlink{0000-0001-5492-4522}$^{2,3}$,
Emanuele Paolo Farina
\orcidlink{0000-0002-6822-2254}$^{4}$,
Eduardo Ba\~{n}ados
\orcidlink{0000-0002-2931-7824}$^{2}$,
\newauthor\
Fabian Walter
\orcidlink{0000-0003-4793-7880}$^{2}$,
Roberto Decarli
\orcidlink{0000-0002-2662-8803}$^{5}$,
Silvia Belladitta
\orcidlink{0000-0003-4747-4484}$^{2,5}$ and
Federica Loiacono
\orcidlink{0000-0002-8858-6784}$^{5}$
\\
$^{1}$Institute of Astronomy, University of Cambridge, Madingley Road, Cambridge, CB3 0HA, UK\\
$^{2}$Max Planck Institut f\"{u}r Astronomie, K\"{o}nigstuhl 17, D-69117, Heidelberg, Germany\\
$^{3}$Department of Astronomy, University of Geneva, Chemin Pegasi 51, 1290 Versoix, Switzerland\\
$^{4}$Gemini Observatory, NSF's NOIRLAB, 670 N A'ohoku Place, Hilo, HI 96720, USA\\
$^{5}$INAF -- Osservatorio di Astrofisica e Scienza dello Spazio di Bologna, Via Gobetti 93/3, 40129 Bologna, Italy
}

\date{Accepted 2024 March 26. Received 2024 March 25; in original form 2024 February 1}

\pubyear{2024}

\begin{document}
\label{firstpage}
\pagerange{\pageref{firstpage}--\pageref{lastpage}}
\maketitle

\begin{abstract}
Of the hundreds of $z\gtrsim6$ quasars discovered to date, only one is known to be gravitationally lensed, despite the high lensing optical depth expected at $z\gtrsim 6$.
High-redshift quasars are typically identified in large-scale surveys by applying strict photometric selection criteria, in particular by imposing non-detections in bands blueward of the $\text{Lyman-}\alpha$ line.
Such procedures by design prohibit the discovery of lensed quasars, as the lensing foreground galaxy would contaminate the photometry of the quasar.
We present a novel quasar selection methodology, applying contrastive learning (an unsupervised machine learning technique) to Dark Energy Survey imaging data.
We describe the use of this technique to train a neural network which isolates an `island' of 11 sources, of which 7 are known $z\sim6$ quasars.
Of the remaining four, three are newly discovered quasars (J0109--5424, $z=6.07$; J0122--4609, $z=5.99$; J0603--3923, $z=5.94$), as confirmed by follow-up Gemini-South/GMOS and archival NTT/EFOSC2 spectroscopy, implying a $91$ per cent efficiency for our novel selection method; the final object on the island is a brown dwarf.
In one case (J0109--5424), emission below the Lyman limit unambiguously indicates the presence of a foreground source, though high-resolution optical/near-infrared imaging is still needed to confirm the quasar's lensed (multiply-imaged) nature.
Detection in the \textit{g} band has led this quasar to escape selection by traditional colour cuts.
Our findings demonstrate that machine learning techniques can thus play a key role in unveiling populations of quasars missed by traditional methods.
\end{abstract}




\section{Introduction
\label{intro}}

High-redshift ($z\gtrsim6$) quasars are important probes of the early Universe, constraining the early growth of supermassive black holes, the properties of the intergalactic medium (IGM), and likely tracing the formation of the first massive structures in the Universe (see e.g.\ \citealt{inayoshi20, volonteri21, fan23} for reviews).

Multi-wavelength large-sky surveys, such as the Dark Energy Survey (DES; \citealt{des}), the VISTA Hemisphere Survey (VHS; \citealt{vhs}), the Wide-field Infrared Survey Explorer (WISE; \citealt{wise}), the Panoramic Survey Telescope and Rapid Response System (Pan-STARRS1; \citealt{panstarrs}) and the Sloan Digital Sky Survey (SDSS; \citealt{sdss}), have enabled the discovery of several hundred high-redshift quasars to date.
Quasars are typically distinguished from other sources, such as stars and galaxies, by their distinctive colour (e.g., \citealt{venemans13, banados15, reed15, reed17}).
The high opacity of the neutral IGM at $z>5.5$ creates a distinctive break below the Lyman-$\alpha$ line, and no flux is expected to be transmitted blueward of the rest-frame Lyman limit ($\lambda=912$ \AA; redshifted to $\lambda>6500$\AA).
Hence traditional $z\gtrsim6$ quasar dropout selection criteria often impose non-detections in the $g$ and $r$ bands (e.g., \citealt{fan01, richards02, reed15, jiang16, Banados2016, reed17, wang19, banados23}).

Such selection criteria are not perfect, and often include contaminants. 
These are most often Galactic brown dwarfs, which have similar colours to high-redshift quasars and are thus difficult to systematically remove (e.g., \citealt{wang19, yang22}).
Moving solar system objects can also contaminate high-redshift quasar searches \citep{bosman23}.
Refinement procedures to remove these contaminants are often time-consuming, and include visual inspection (e.g., \citealt{venemans13, banados15, reed15}), SED fitting (e.g., \citealt{reed17, andika23}), or extreme deconvolution (e.g., \citealt{riccardo22}).

Magnification of high-redshift quasars due to gravitational lensing by a galaxy on the same line of sight enables the study of quasars of intrinsically lower luminosity, and yields insights into their accretion discs (e.g., \citealt{chan21}) and host galaxies (e.g., \citealt{stacey18}).
Observations of lensed quasars also place constraints on the dark matter profiles in the lensing galaxies (e.g., \citealt{gilman20}), and time-delay cosmography can be used to estimate the Hubble constant independently of the discrepant early- and late-Universe measurements \citep{refsdal64, treu22}.

However, it has long been known that traditional photometric selection criteria exclude lensed high-redshift quasars, as flux from the galaxy can pollute the quasar colour to the extent that it escapes selection (e.g.\ \citealt{wyithe02}, \citealt{pacucci19}).
Indeed, the first (and to date, only confirmed) strongly-lensed $z>6$ quasar to be discovered, J0439+1634 at $z=6.51$, had such a faint lensing galaxy that if the lens were only $0.5\,\text{mag}$ brighter, the quasar would not have met the selection criteria applied \citep{fan19}.
As a result, the fraction of high-redshift quasars observed to be lensed is significantly lower than expected.
Indeed, the high lensing optical depth at $z\gtrsim6$ means that the lensed fraction is likely to be higher, with various studies suggesting it should be at least about 1 per cent \citep{yue22} or as high as 20 per cent \citep{pacucci19}.
It is therefore likely that the selection criteria used to identify high-redshift quasars are excluding an entire population.

This paper is not the first attempt to select lensed quasars. \citet{andikaensemble} use an ensemble of state-of-the-art convolutional neural networks (CNNs) trained on mock images to identify 3080 candidate lensed quasars at $z>1.5$ in the Hyper Suprime-Cam Subaru Strategic Survey (HSC-SSP; \citealt{hscssp}), pared down by visual inspection and astrometric information to 210, awaiting spectroscopic confirmation.
At $z\gtrsim6$, \citet{andika23} use a combination of CNNs and SED fitting to identify 448 lensed and unlensed candidate high-redshift quasars in DES, reduced by visual inspection to 36, which also await confirmation.
\citet{yue23} train a probabilistic random forest for their colour selection criteria, followed by morphological selection and a visual inspection phase, discovering an intermediately-lensed (magnified but not multiply-imaged) quasar, as well as a quasar pair, at $z\sim 5$.

The above are just three examples of the numerous and varied applications which machine learning (ML) is finding in the natural sciences.
However, the reliance of many ML techniques on accurately labelled training data often makes such methods impractical.
In this specific case, the absence of a large training set of real lensed $z\sim6$ quasars is a key weakness of supervised methods.
Unsupervised ML circumvents this by comparing instances to each other and grouping them according to similarity.
In particular, contrastive learning (e.g.\ \citealt{chen20}) seeks a low-dimensional representation of the data by training a CNN to project innocuous transformations (rotations, reflections, etc.) of the same image closer together in a latent space, while aggressively separating projections of different sources.

This paper presents a new contrastive-learning-based methodology to find high-redshift quasars, lensed or not.
We demonstrate the efficiency of this new method in isolating an `island' of 11 sources, of which 10 are quasars.
This includes three new high-redshift quasars confirmed spectroscopically, one of which may have been missed by previous searches due to its detection in the \textit{g} band.
One source spectrum exhibits flux below the Lyman limit, unambiguously indicating the presence of a foreground source which may be lensing the background quasar.
This paper is structured as follows. 
In Section \ref{methods} we describe our candidate selection criteria, the contrastive learning approach, and SED fitting of our targets.
Section \ref{confirmation} presents the observations confirming these targets as high-redshift quasars.
We discuss implications of these observations in Section~\ref{discussion}, and summarize our work in Section \ref{conc}.

Magnitudes reported in this paper are all in the AB system.
Where required, we use a flat cosmology with $H_0=70\,\text{km}\,\text{s}^{-1}\,\text{Mpc}^{-1}$ and $\Omega_{\mathrm{m}0}=0.3$.

\section{Methods \& Target Selection} \label{methods}

\begin{figure}
\includegraphics[width=\columnwidth]{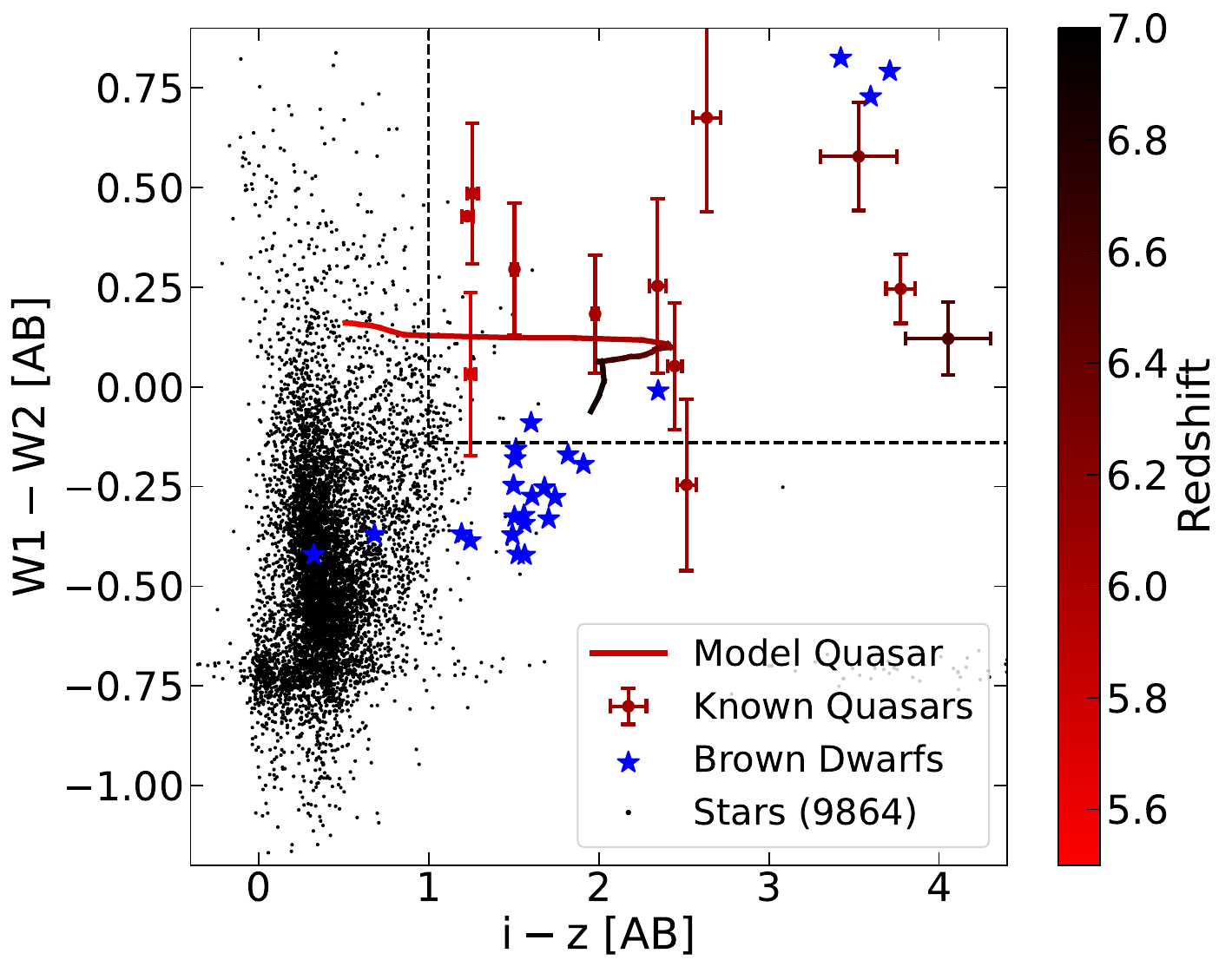}
\vspace{-.4cm}
\caption{Photometric selection criteria.
The black points are sources selected from three random DES tiles, most of which are likely stars.
The blue stars indicate brown dwarfs from \citet{kirkpatrick11} and \citet{best15} cross-matched to DES.
Several known high-redshift quasars are plotted, as well as a model quasar track generated using \textsc{QSOGEN} \citep{temple22}; both are coloured by redshift.
Our liberal photometric selection criteria are shown by the dashed lines: $\mathit{i}-\mathit{z}>1$ and $\mathit{W1}-\mathit{W2} > -0.14$ (both AB).
}
\label{fig:colours}
\end{figure}

\begin{figure*}
\includegraphics[width=0.8\textwidth]{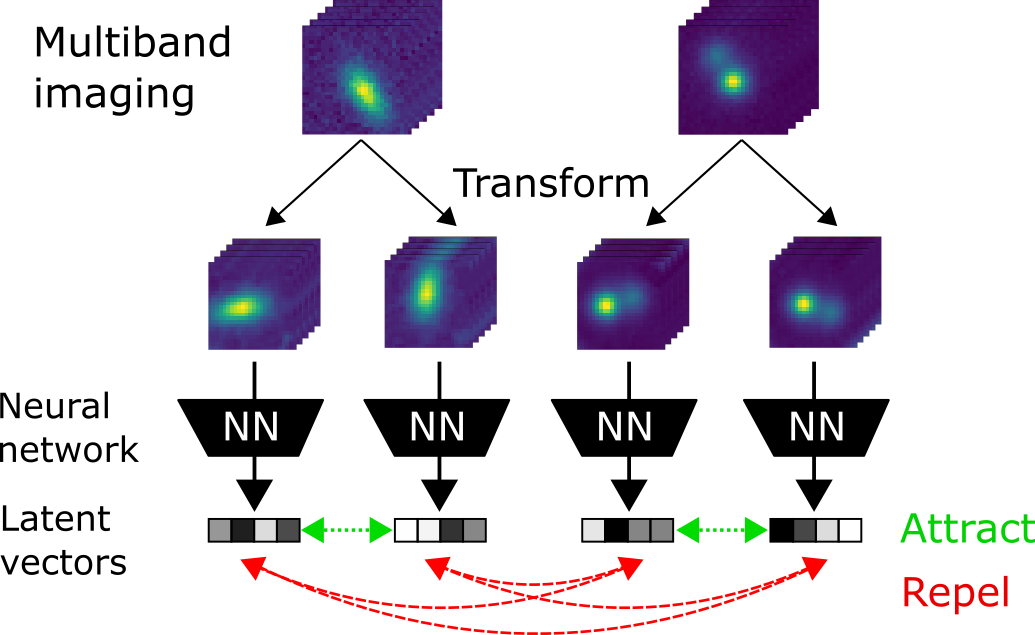}
\caption{Schematic of contrastive learning.
A batch of $N$ \textit{grizy} images are each randomly transformed twice, before being inputted into a convolutional neural network.
The output of this network is a set of $2N$ vectors in a latent space.
The contrastive loss function, computed using the locations of these vectors in the latent space, encourages proximity between vectors from the same source (the green dotted ``attract'' arrows), and discourages proximity between different sources (red dashed ``repel'' arrows).
The training procedure is illustrated above for a batch size of 2; in practice a larger batch size is used (we use 128, see Table~\ref{tab:hyperparameters}).
The high-dimensional latent vectors output by the neural network are represented by the shaded sets of 4 boxes; in practice the latent dimension is larger (we use 128).
}
\label{fig:contrastive}
\end{figure*}

\subsection{Data preparation} \label{dataprep}

We have used photometric and imaging data from the DES Data Release 2 (DR2), which covers 5000 deg$^2$ of the southern celestial hemisphere in the broad \textit{grizY} bands \citep{desdr2}.
The DES DR2 catalogue contains photometry and imaging for almost 700 million objects.
Here we outline the criteria used to select our working sample;
some of these criteria are also shown in Fig.~\ref{fig:colours}.
Our photometric selection criteria are less stringent than in previous studies in order to accommodate lensed quasars, where the lensing galaxy is expected to alter the observed quasar colour.
All \textit{grizY} magnitudes quoted in this work use the DES DR2 aperture 4, equivalent to $1.95\,\text{arcsec}$ diameter \citep{desdr2}.

We first select sources with $\mathit{z} < 21.0$ and $\mathit{Y}<22.45$; we also impose magnitude errors of $<0.1$ on each of these bands.
We further select sources with $\mathit{i}-\mathit{z}>1.0$ to preferentially select quasars and remove contaminants such as stars and galaxies, though this is a lower cutoff than in many previous searches (e.g.\ \citealt{reed15, reed17, wang19, Banados2016, banados23}) to allow for photometric contamination from a lensing galaxy.
Unlike most previous work, no requirement is placed on the \textit{g} or \textit{r} bands.
We apply further standard cuts on the \texttt{IMAFLAGS\_ISO} and \texttt{FLAGS} parameters to remove artefacts such as cosmic rays and saturation trails \citep{desdr2}.
The DES DR2 SQL query combining these criteria can be found in Appendix \ref{app_sql}.
These selection criteria result in 218\,241 objects.

We further require a counterpart to each DES source in both VHS DR5 \citep{vhsdr5} and AllWISE \citep{allwise} within $1\,\text{arcsec}$, to remove contaminants as well as artefacts that would be present in only one survey. 
After cross-matching to both surveys, 116\,499 objects remain.
We then use a colour cut of $\mathit{W1}-\mathit{W2}>-0.14 \,\text{mag [AB]}$ (corresponding to $\mathit{W1} - \mathit{W2} > 0.5 \,\text{mag [Vega]}$), to remove a large number of cool dwarf stars, whose \textit{grizY} photometry is often similar to that of high-redshift quasars (e.g.\ \citealt{carnall15}).
After further imposing detection in \textit{J}, \textit{K}, \textit{W1} and \textit{W2} bands, with S$/\text{N}>2$ in all bands, there are 7\,438 objects remaining.

Despite the large number of remaining objects, no further colour cuts are made, to prevent lensed quasars from being excluded.
Traditional photometric selection criteria are usually more restrictive, requiring higher thresholds for the $\mathit{i}-\mathit{z}$ colour, making extra cuts on the $\mathit{z}-\mathit{Y}$ or $\mathit{Y}-\mathit{J}$ colours, and requiring non-detections in the \textit{g} and \textit{r} bands (e.g., \citealt{banados14, Banados2016}).
Some also take steps to exclude extended sources at this stage (e.g., \citealt{reed15}), but we delegate morphology-based selection to our ML procedure (see Section~\ref{contrastive}).
Stricter criteria would naturally lead to a significantly smaller sample size ($\sim 10^2$ objects in the DES--VHS--WISE footprint, e.g.\ \citealt{reed15, reed17}) from which it is not impractical to simply inspect the sources individually or run SED-fitting procedures to select good quasar candidates.
However, one of the aims of this work is to dispense with visual inspection altogether.

Furthermore, given the relatively low number density of high-redshift quasars (e.g.\ \citealt{schindler23}) it is likely that our sample of 7\,438 objects is dominated by contaminants, even taking into account a potentially large fraction of lensed quasars at $z\sim 6$ (e.g., \citealt{pacucci19, yue22}).
We suggest that it is difficult and perhaps impossible to remove these contaminants by any process based solely on photometry without also removing lensed high-redshift quasars: it is likely that they can occupy the same regions of colour space.
Indeed, Fig.~\ref{fig:colours} shows that the $\mathit{i}-\mathit{z}$ and $\mathit{W1}-\mathit{W2}$ colours of high-redshift quasars are shared by brown dwarfs and other objects (such as dusty low-redshift galaxies).

Instead, we suggest that it is possible in principle to distinguish high-redshift quasars from contaminants using source imaging, another data product of DES \citep{des}.
The large number of candidates motivates the use of an ML-based selection, particularly as the application of ML to image classification has rapidly improved in recent years (e.g.\ \citealt{lecun15, pak17}).

\subsection{Contrastive Learning} \label{contrastive}

A requirement of supervised ML techniques is the availability of accurate ground-truth labels for the data: poorly-labelled data can severely impede image classification \citep{frenay14}.
Although several known high-redshift quasars are present in our sample, it is \textit{a priori} unknown whether lensed quasars would appear substantially different in DES imaging.
The only known lensed high-redshift quasar (J0439+1634) has been described in detail \citep{fan19}, but is not in the DES footprint.
Additionally, the serendipity of its discovery and its high magnification ($\mu\sim50$) suggest that its appearance could be unrepresentative of the lensed high-redshift quasar population:
magnification distributions are typically taken to be $\propto \mu^{-3}$ \citep{schneider92, wyithe11, yue22}.
It would therefore be inappropriate to label objects as lensed high-redshift quasars based on subjective similarity to J0439+1634.
In any case, it would be impractical to individually inspect and attempt to label each image in our large working sample.

Unsupervised ML techniques do not require ground-truth data labels, circumventing the above issues.
Not only does this remove the necessity to label each image by hand, it avoids biases due to any prior assumptions about the appearance of lensed quasars in DES imaging.

The novelty in our candidate selection procedure is the use of contrastive learning, an unsupervised ML technique \citep{chen20}.
Briefly, this method trains a neural network to group together similar images in a high-dimensional latent space.
The loss function used to train the network encourages proximity in the latent space between projections of transformations of the \textit{same} images, and penalizes proximity of transformations of \textit{different} images.
Over the course of the self-supervised training, similar objects thus end up being clustered together in the latent space.
A schematic of this technique is shown in Fig.~\ref{fig:contrastive} and described below; the training hyperparameters are given in Table~\ref{tab:hyperparameters} in Appendix~\ref{app_contrastive}.

The contrastive learning procedure is as follows.
A batch of $N$ images is selected at random from the data; each ``image" consists of $28\times28$ pixel cutouts in the 5 \textit{grizY} bands of DES ($7.5\times7.5\arcsec$), normalized to the brightest pixel across all five bands; we use here the fact that the photometric calibrations of the DES images are uniform \citep{burke18}.
Two sequences of random transformations are then made of each image, before projection with a CNN.
These transformations are such that they should not affect the classification of the image: it is cropped and may be flipped, rotated, and translated, but the relative brightness of the source in the different bands -- that is, the colour -- is not changed, as this can be a diagnostic feature.
Similarly, the image is not sheared or zoomed in any way, as this could (for example) transform an image of a quasar into an image resembling an extended galaxy.
This transformation stage ensures that the learning process uses only information relevant to source classification, such as colour, shape, and extension; not e.g.\ orientation, centring.

These $2N$ transformed images are then each inputted into a CNN, whose architecture is inspired by those of \citet{sarmiento21} and \citet{chen20}, and outlined in Table~\ref{tab:architecture} in Appendix~\ref{app_contrastive}.
Contrastive learning models consist of three modules: data augmentation, a base encoder, and a projection head.
The data augmentation module transforms each image as discussed above:
randomly cropping to $24\times24\,\text{px}$;
randomly flipping it horizontally and/or vertically;
randomly translating it by $2\,\text{px}$;
randomly rotating it.
For a given object, an identical transformation is applied to the images in each band.
The base encoder consists of a sequence of three 2D convolutional layers \citep{fukushima79, lecun15} with exponential linear unit activation \citep{elu}; each layer is followed by a max-pooling layer with pool size 2 \citep{weng92}.
The output of this module is reshaped to a 512-dimensional vector.
Finally, the projection head consists of three densely-connected layers, outputting a vector $\vb*{z}_i$ in a high-dimensional (in this case 128-dimensional) latent space.

The training process consists of altering the weights of the neural network so that similar images are projected to closer vectors in the latent space, and different images are projected further away from each other.
The contribution of each pair of representations to the loss function is
\begin{equation}
L_{i;j} = -\log \frac{
    \exp(
        \hat{\vb*{z}}_i \cdot \hat{\vb*{z}}_j
        / \tau
    )
}{
    \sum_{k=1, k\neq i}^{2N}
        \exp(
            \hat{\vb*{z}}_i \cdot \hat{\vb*{z}}_k
            / \tau
        )
},
\label{eq:loss}
\end{equation}
where $\hat{\vb*{z}}_i = \vb*{z}_i / \lVert\vb*{z}_i\rVert_\text{L2}$, the vector $\vb*{z}_j$ is that which came from the same original DES object as the vector $\vb*{z}_i$, and $\tau$ is a `temperature' hyperparameter affecting the clustering of the representations.
The loss for a batch is then the average of the above contributions over all $2N$ output vectors.
In minimizing the loss function over multiple batches (and hence multiple different stochastic transformations), the network thus learns to group together latent vectors from different transformations of the similar sources, while separating all others.

Following training, the projection head is discarded and the images are inputted directly to the base encoder;
\citet{chen20} showed that removing the projection head improves performance by over 10 per cent.
The output of the encoder is a 512-dimensional vector for each object.
We then apply dimensionality reduction to this set of vectors using $t$-distributed Stochastic Neighbour Embedding ($t$-SNE; \citealt{tSNE}), to be able to visualize the latent space in two dimensions while approximately preserving the pairwise distances between the vectors.

\begin{figure*}
\includegraphics[width=0.75\textwidth]{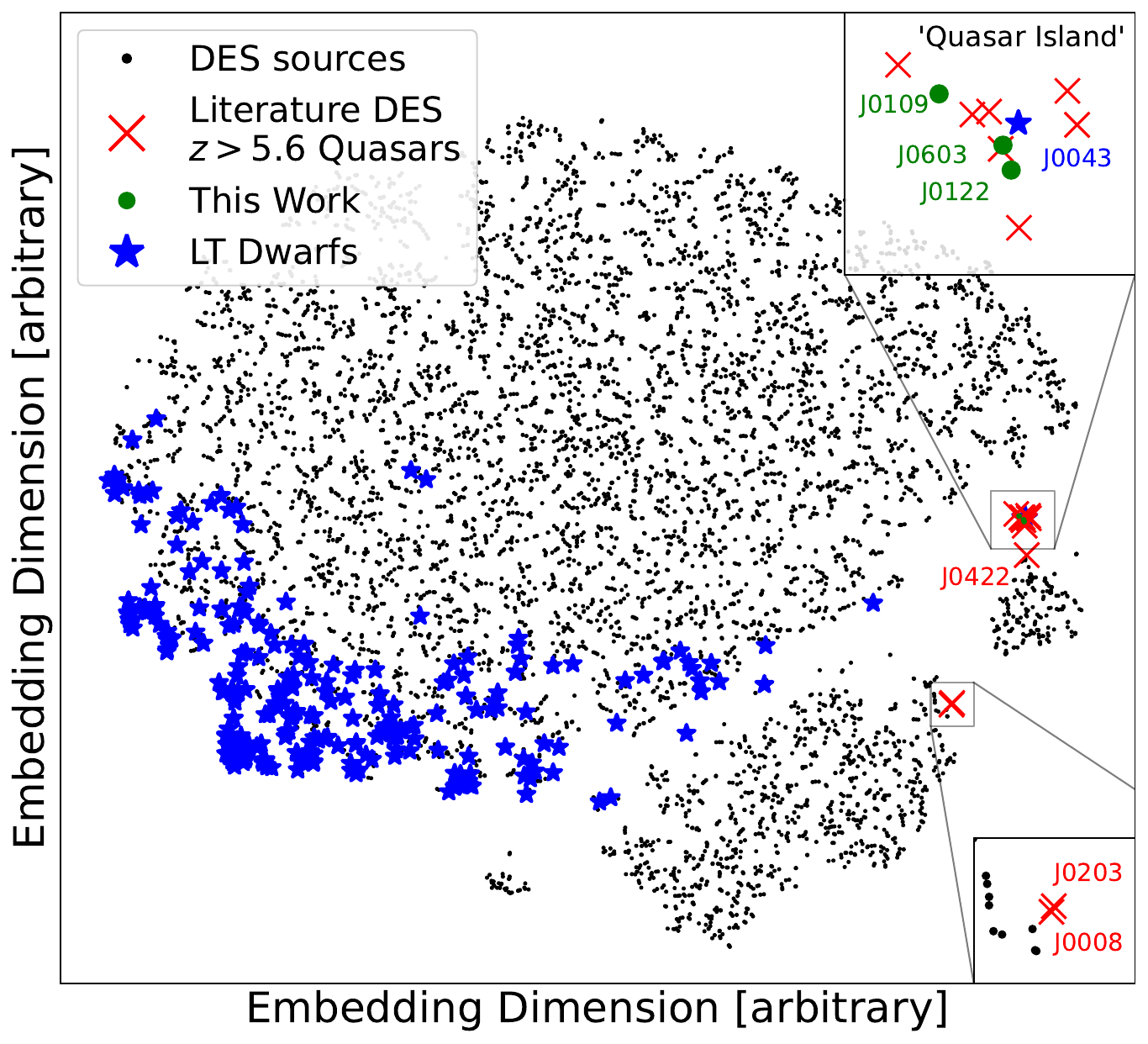}\\
\caption{
Dimensionality-reduced projection of DES images.
The $t$-SNE procedure reduces the dimensionality of the set of vectors output by the encoder of the CNN while attempting to preserve the relative distances between each pair of points
(as such the locations of the points would be equally valid if the plot were arbitrarily re-scaled or rotated).
There are several distinct groups of objects, including an isolated island towards the right of the frame, which we call `Quasar Island' and show in the top right inset.
This island contains 11 objects, of which seven are known high-redshift quasars (red crosses), three are newly-discovered quasars (green circles), and one is a brown dwarf (blue star).
Three other literature high-redshift quasars are peripheral members of other islands: the one projected just below `Quasar Island' and two in the lower right inset.
The blue stars mostly at the lower left edge of the main continent are LT dwarf candidates from \citet{dalponte23}.
The three other islands towards the bottom and right of the frame contain sources with extended morphologies, likely galaxies.
}
\label{fig:latent}
\end{figure*}

\subsection{Identification of DES quasar candidates}

The result of the unsupervised clustering and dimensionality reduction is shown in Fig.~\ref{fig:latent}.
The neural network, trained using contrastive learning as described above, has separated the sample defined in Section~\ref{dataprep} into several defined clusters.
An `island' towards the right of the frame contains 11 objects, including 7 known high-redshift quasars, which we term `Quasar Island'.
The location of this island was not known \textit{a priori}, but is simply defined as the most significant cluster of known high-redshift quasars.
Follow-up spectroscopy of the other four objects (see Section~\ref{confirmation}) showed three to be newly discovered high-redshift quasars, and one to be a cool dwarf.
We summarize the properties of these four objects in Tables~\ref{tab:quasarisland} and \ref{tab:quasarislandmags}.
One of these sources (J0109--5424) was likely discarded by previous searches due to its detection in the \textit{g} band; this detection indicates contamination by a foreground object (see Section~\ref{sec:lensed_quasar}).
The fact that the neural network identified this source as being similar to other high-redshift quasars, despite being detected in the \textit{g} band, suggests that (dis)similarity in the other four bands is a more efficient metric to cluster the objects considered by the network. We therefore hypothesize that a pre-selection of sources (e.g. using the colour cuts detailed above) is an important step to ensure the success of our selection method.

Determining the detailed nature of the other clusters is beyond the scope of this work, but we briefly mention their main characteristics here.
By inspection of a few candidates, the largest `continent' appears to contain isolated, compact sources: most of these objects are likely to be cool dwarfs; the lower left edge of this continent contains many L and T dwarf candidates from \citet{dalponte23}.
The three other islands towards the bottom and right of the frame contain sources with extended morphologies, indicating that they are galaxies. They also appear to be separated according to distinct \textit{grizY} colours.
We note that three peripheral members of these other islands are known high-redshift quasars: the one just below the quasar island (J0422--1927, \citealt{Banados2016}), and two in the lower right inset (J0203+0012, \citealt{venemans07}; J0008--0626, \citealt{Banados2016}).
This highlights the fact that this method does not identify all the quasars in the sample, so there may also be some unidentified quasars that have been missed by our method.

\begin{table*}
\centering
\caption{
Coordinates, redshifts, and absolute magnitudes of the 11 objects clustered together on `Quasar Island', and the three known high-redshift quasars not on this island.
The first four are investigated by this work, with the neural network suggesting similarity between these and the seven other high-redshift quasars discovered by other authors.
Redshifts are obtained spectroscopically, except for J0043--6028 which is a cool dwarf.
The absolute magnitude at $1450\,\text{\AA}$ is computed using the best-fitting quasar spectrum (see Fig. \ref{fig:fits}) and the \citet{temple22} template rest-frame UV slope $\alpha_\nu=-0.341$.
}
\label{tab:quasarisland}
\begin{tabular}{cccccc}
\hline
Object & R.A. & Dec. & Redshift & $M_{1450}$ & Discovery \\
\hline
\textbf{J0109--5424} & $01^\text{h}09^\text{m}09\fs00$ & $-54\degr24\arcmin16\farcs29$ & 6.074 & $-26.52$ & This work \\
\textbf{J0603--3923} & $06^\text{h}03^\text{m}52\fs26$ & $-39\degr23\arcmin35\farcs78$ & 5.941 & $-25.83$ & This work \\
\textbf{J0122--4609} & $01^\text{h}22^\text{m}57\fs70$ & $-46\degr09\arcmin14\farcs20$ & 5.986 & $-26.51$ & This work \\
\textit{\textbf{J0043--6028}} & $00^\text{h}43^\text{m}59\fs67$ & $-60\degr28\arcmin44\farcs85$ & - & - & This work \\
ATLAS~J029.9915--36.5658 & $01^\text{h}59^\text{m}57\fs98$ & $-36\degr33\arcmin56\farcs67$ & 6.020 & $-27.0$ & \citet{carnall15}\\
VDESJ0454--4448 & $04^\text{h}54^\text{m}01\fs79$ & $-44\degr48\arcmin31\farcs05$ & 6.100 & $-26.5$ & \citet{reed15} \\
PSO~J056.7168--16.4769 & $03^\text{h}46^\text{m}52\fs05$ & $-16\degr28\arcmin36\farcs98$ & 5.990 & $-26.72$ & \citet{Banados2016} \\
VDESJ0408--5632 & $04^\text{h}08^\text{m}19\fs24$ & $-56\degr32\arcmin28\farcs76$ & 6.035 & $-26.51$ & \citet{reed17} \\
VDESJ0224--4711 & $02^\text{h}24^\text{m}26\fs55$ & $-47\degr11\arcmin29\farcs36$ & 6.526 & $-26.93$ & \citet{reed17} \\
VDESJ0143--5545 & $01^\text{h}43^\text{m}10\fs25$ & $-55\degr45\arcmin10\farcs71$ & 6.250 & $-25.65$ & \citet{reed17} \\
PSO032.91882--17.07465 & $02^\text{h}11^\text{m}40\fs53$ & $-17\degr04\arcmin28\farcs77$ & 6.050 & $-25.80$ & \citet{banados23} \\
\hline
PSO~J065.5041--19.4579 & $04^\text{h}22^\text{m}01\fs00$ & $-19\degr27\arcmin28\farcs69$ & 6.125 & $-26.62$ & \citet{Banados2016}\\
ULAS~J0203+0012 & $02^\text{h}03^\text{m}32\fs38$ & $+00\degr12\arcmin29\farcs06$ & 5.709 & $-26.2$ & \citet{venemans07} \\
PSO~J002.1073--06.4345 & $00^\text{h}08^\text{m}25\fs77$ & $-06\degr26\arcmin04\farcs42$ & 5.929 & $-26.0$ & \citet{Banados2016} \\
\hline
\end{tabular}
\end{table*}

\begin{table*}
\caption{
DES DR2 \texttt{MAG\_APER\_4}, VHS and WISE magnitudes of the 11 objects clustered together on `Quasar Island', and the three known high-redshift quasars not on this island.
All magnitudes are in the AB system.
Upper limits are given at the $2\sigma$ level.
} \label{tab:quasarislandmags}
\scalebox{0.85}{
\begin{tabular}{cccccccccc}
\hline
Name & \textit{g} & \textit{r} & \textit{i} & \textit{z} & \textit{Y} & \textit{J} & \textit{K} & \textit{W1} & \textit{W2} \\
\hline
\textbf{J0109--5424} & $25.07 \pm 0.15$ & $23.09 \pm 0.03$ & $22.73 \pm 0.04$ & $20.18 \pm 0.01$ & $20.19 \pm 0.03$ & $19.53 \pm 0.11$ & $19.01 \pm 0.15$ & $19.12 \pm 0.06$ & $18.72 \pm 0.08$ \\
\textbf{J0603--3923} & $>26.45$ & $25.34 \pm 0.19$ & $22.43 \pm 0.02$ & $20.78 \pm 0.01$ & $20.85 \pm 0.04$ & $20.10 \pm 0.16$ & $20.08 \pm 0.42$ & $19.83 \pm 0.09$ & $19.64 \pm 0.14$ \\
\textbf{J0122--4609} & $>26.45$ & $24.28 \pm 0.10$ & $21.89 \pm 0.02$ & $20.09 \pm 0.01$ & $20.28 \pm 0.03$ & $19.73 \pm 0.18$ & $19.39 \pm 0.24$ & $19.41 \pm 0.08$ & $19.40 \pm 0.14$ \\
\textit{\textbf{J0043--6028}} & $>26.45$ & $24.86 \pm 0.16$ & $22.15 \pm 0.03$ & $20.49 \pm 0.01$ & $20.52 \pm 0.04$ & $19.49 \pm 0.06$ & $20.02 \pm 0.44$ & $20.49 \pm 0.18$ & $19.84 \pm 0.19$ \\
ATLAS~J029.9915--36.5658 & $>26.45$ & $24.03 \pm 0.07$ & $21.86 \pm 0.02$ & $19.88 \pm 0.01$ & $20.03 \pm 0.02$ & $19.15 \pm 0.10$ & $19.17 \pm 0.21$ & $19.40 \pm 0.08$ & $19.21 \pm 0.12$ \\
VDESJ0454--4448 & $>26.45$ & $25.51 \pm 0.26$ & $22.93 \pm 0.04$ & $20.49 \pm 0.01$ & $20.56 \pm 0.03$ & $20.32 \pm 0.14$ & $20.41 \pm 0.46$ & $19.68 \pm 0.07$ & $19.62 \pm 0.14$ \\
PSO~J056.7168--16.4769 & $>26.45$ & $24.27 \pm 0.09$ & $21.66 \pm 0.02$ & $20.15 \pm 0.01$ & $20.52 \pm 0.04$ & $19.89 \pm 0.18$ & $19.48 \pm 0.30$ & $19.50 \pm 0.09$ & $19.20 \pm 0.14$ \\
VDESJ0408--5632 & $>26.45$ & $25.50 \pm 0.33$ & $22.69 \pm 0.05$ & $20.34 \pm 0.01$ & $20.38 \pm 0.03$ & $19.92 \pm 0.13$ & $19.48 \pm 0.26$ & $20.43 \pm 0.12$ & $20.18 \pm 0.18$ \\
VDESJ0224--4711 & $>26.45$ & $>26.15$ & $24.56 \pm 0.25$ & $20.51 \pm 0.01$ & $20.37 \pm 0.03$ & $19.88 \pm 0.14$ & $19.00 \pm 0.11$ & $18.88 \pm 0.05$ & $18.76 \pm 0.08$ \\
VDESJ0143--5545 & $>26.45$ & $>26.15$ & $24.38 \pm 0.23$ & $20.85 \pm 0.02$ & $21.28 \pm 0.07$ & $20.77 \pm 0.27$ & $19.77 \pm 0.29$ & $19.65 \pm 0.09$ & $19.07 \pm 0.10$ \\
PSO032.91882--17.07465 & $>26.45$ & $>26.15$ & $23.61 \pm 0.08$ & $20.97 \pm 0.02$ & $21.08 \pm 0.07$ & $20.34 \pm 0.16$ & $19.40 \pm 0.23$ & $20.06 \pm 0.16$ & $19.39 \pm 0.17$ \\
\hline
PSO~J065.5041--19.4579 & $>26.45$ & $>26.15$ & $23.66 \pm 0.09$ & $19.88 \pm 0.01$ & $20.79 \pm 0.04$ & $19.83 \pm 0.08$ & $19.10 \pm 0.15$ & $18.67 \pm 0.05$ & $18.42 \pm 0.07$ \\
ULAS~J0203+0012 & $>26.45$ & $25.44 \pm 0.29$ & $22.16 \pm 0.03$ & $20.92 \pm 0.01$ & $21.11 \pm 0.07$ & $19.76 \pm 0.14$ & $19.23 \pm 0.17$ & $19.39 \pm 0.09$ & $19.36 \pm 0.19$ \\
PSO~J002.1073--06.4345 & $>26.45$ & $24.44 \pm 0.16$ & $21.98 \pm 0.03$ & $20.72 \pm 0.02$ & $21.05 \pm 0.08$ & $20.36 \pm 0.11$ & $19.83 \pm 0.25$ & $19.51 \pm 0.11$ & $19.02 \pm 0.14$ \\
\hline
\end{tabular}
}
\end{table*}

\section{Follow-up confirmation and archival search} \label{confirmation}

\subsection{An archival \texorpdfstring{$z\sim 6$}{TEXT} quasar: J0122--4609}
\label{archival}
\begin{figure*}
    \centering
    \includegraphics[width=0.9\textwidth]{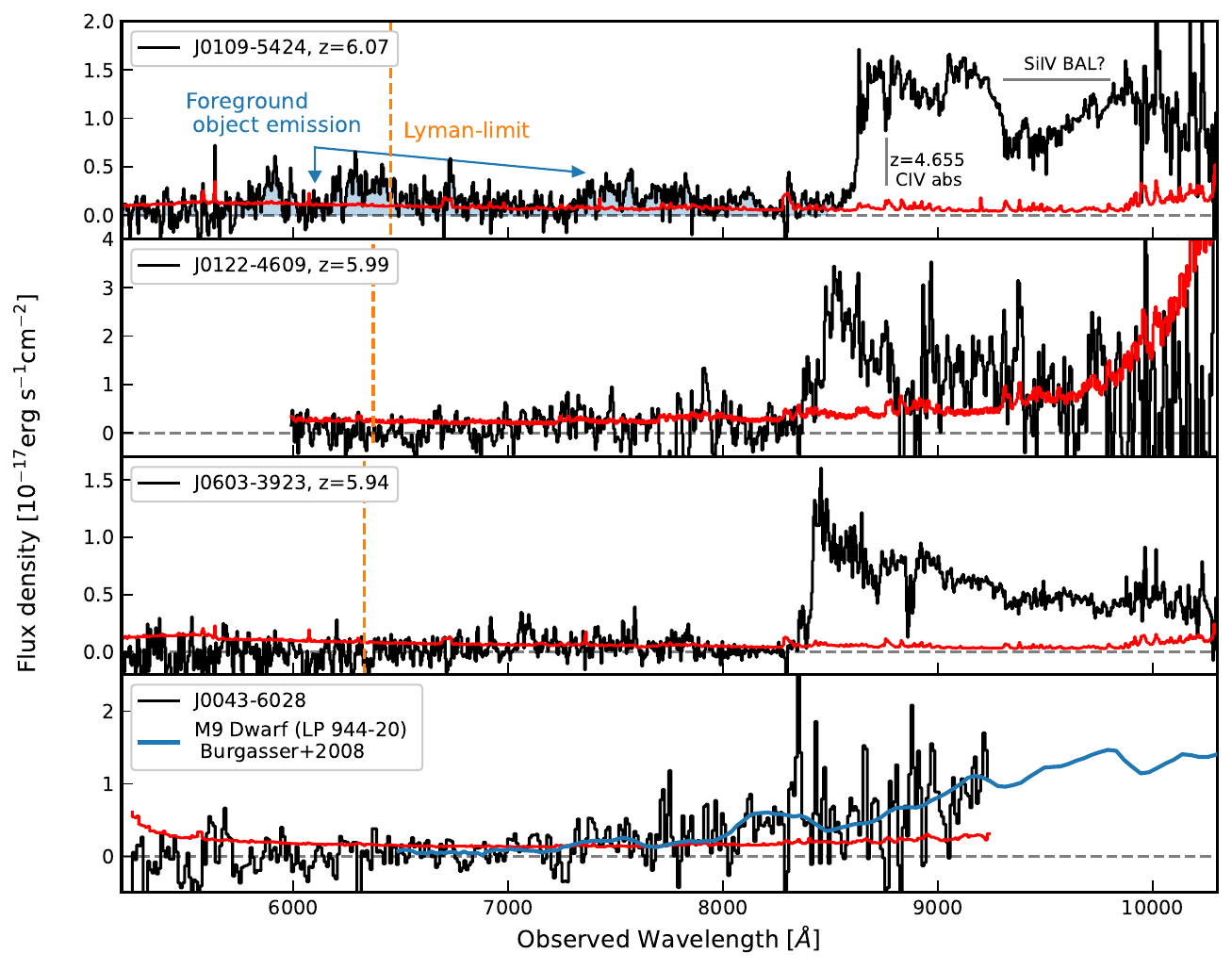}
    \caption{
    Confirmation spectra of the four unpublished objects of `Quasar Island' studied in this work with EFOSC2/NTT and Gemini-South/GMOS.
    The spectra are all flux-calibrated and scaled to the respective DES \textit{z}-band magnitudes (see Table~\ref{tab:quasarislandmags}).
    We show the extracted 1D error array in red.
    We confirm the discovery of three new quasars at $z\sim 6$, bringing the completeness of `Quasar Island' to $91$ per cent.
    The remaining object is a cool dwarf whose spectrum closely matches that of an M9 dwarf \citep{Burgasser2008}.
    For J0109--5424, we highlight in blue the presence of flux in the Lyman-$\alpha$ forest and below the Lyman limit (orange dashed line), which we interpret as evidence for a foreground lensing galaxy (see Section~\ref{sec:lensed_quasar}); we also identify possible \ion{C}{IV} and \ion{Si}{IV} absorptions.
    For the other two quasars, no flux is detected below the Lyman limit. }
    \label{fig:all_spectra}
\end{figure*}

We cross-matched the four unconfirmed candidates with the ESO archive and Gemini-South archive.
We found that J0122--4609 had unpublished EFOSC2/NTT spectroscopy (programme 098.A-0439(A), PI: McMahon).
It was observed on 2016 December 22/23 for $1200\,\text{s}$ using the $1.0\,\text{arcsec}$-wide slit with grism $\#16$ and the OG530 blocking filter.
The spectrum was reduced with the \textsc{Pypeit} package \citep{pypeit:joss_pub}, using the associated bias, flats and standard returned by the ESO archive.
The final spectrum is flux-calibrated and re-scaled to match the DES \textit{z}-band magnitude (see Table \ref{tab:quasarislandmags}).
The spectrum clearly shows the Lyman break and blue rest-frame UV colour of a high-redshift quasar (see Fig.~\ref{fig:all_spectra}, second panel).
We fit the spectrum of J0122--4609 with the `median' quasar template of \citet{Banados2016}, deriving a redshift of $z=5.99$.

\subsection{Discovery of two new quasars at \texorpdfstring{$z\sim 6$}{TEXT}, including a lensed candidate}

Two further candidates on `Quasar Island', J0603--3923 and J0109--5424, were observed on 2022 December 6 and 11 respectively with the Gemini Multi-Object Spectrograph (GMOS) South (Programme ID GS-2022B-FT-208, PI: Farina).
The two candidates were observed in long-slit mode for $4\times300\,\text{s}$ exposures using the $1.5\,\text{arcsec}$ slit width.
We used the R400+G5325 grating in combination with the GG455\_G0329 blocking filter to cover the wavelength range $5300\,\text{\AA}\leq \lambda \leq 10300\,\text{\AA}$ where the Lyman break of a $z\sim 6$ quasar would be expected.
As with J0122--4609 above, the spectra were reduced using \textsc{Pypeit} \citep{pypeit:joss_pub}, and the final flux-calibrated spectrum is re-scaled to match the DES \textit{z}-band magnitude (see Table \ref{tab:quasarislandmags}).

We show the optical GMOS spectra of these two sources in Fig.~\ref{fig:all_spectra}, revealing a clear Lyman-$\alpha$ break typical of $z\sim 6$ quasars. Notably, J0109--5424 shows an excess of flux in the Lyman-$\alpha$ forest and below the Lyman continuum limit, consistent with the \textit{g} and \textit{r} band detection which we interpret as evidence for a foreground lensing galaxy (see further Section~\ref{sec:lensed_quasar}).
We fit J0603--3923 with the `median' quasar template of \citet{Banados2016} and J0109--5424 with the `weak' template, deriving best-fitting redshifts of $z=5.94$ and $6.07$, respectively. 
The error of these templates compared to the \ion{Mg}{ii} or [\ion{C}{ii}] redshifts is typically $\Delta z<0.03$ \citep{banados23}.
Both quasars are detected in the \textit{r} band, and J0109--5424 also in the \textit{g} band, which may have led them to be discarded as contaminants in traditional colour-cut searches.
However, the GMOS spectra confirm their quasar nature, validating the efficiency of our contrastive network to find lensed quasar candidates.

\begin{figure*}
\includegraphics[width=\textwidth]{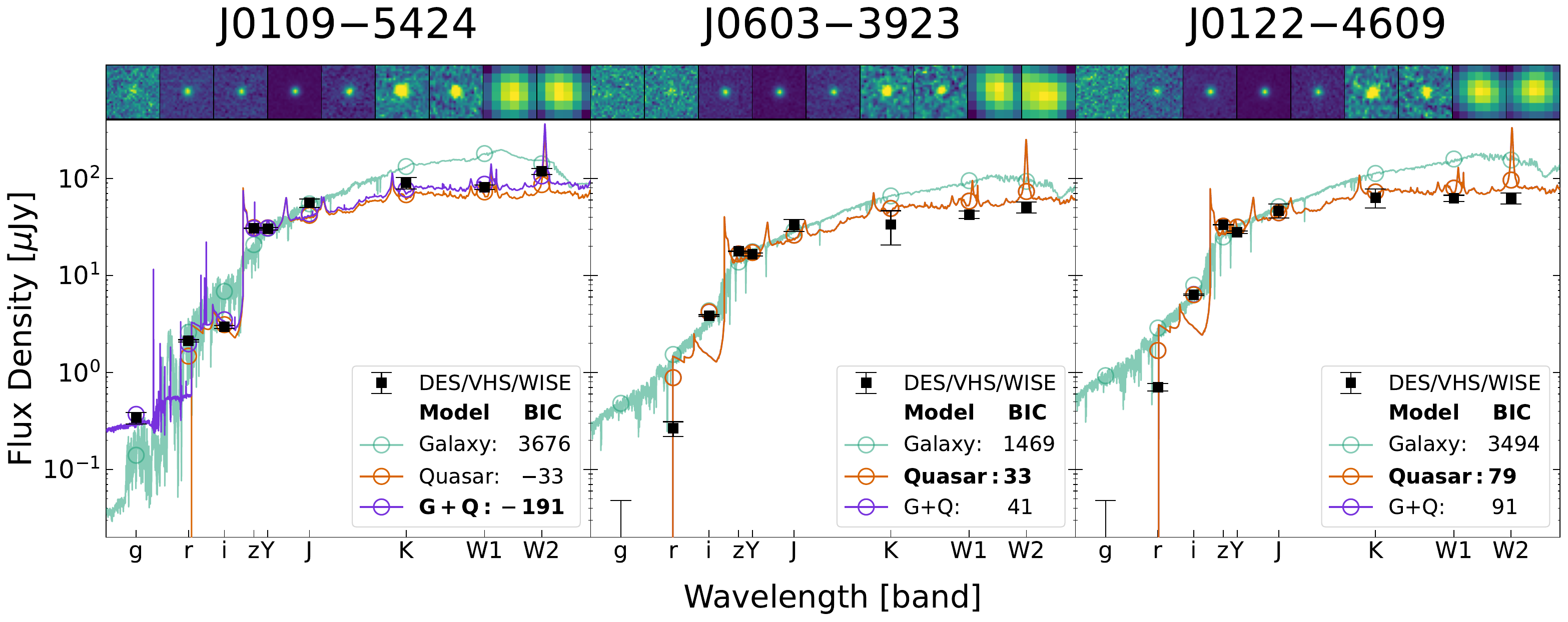}
\caption{
DES+VHS+WISE imaging and best-fitting spectra of the candidates' photometry for galaxy, quasar and lensed quasar templates based on \textsc{BAGPIPES} galaxy models \citep{carnall18} and \textsc{QSOGEN} quasar models \citep{Temple2021,temple22} as described in Section~\ref{sec:lensed_quasar}.
The images are each $7.5\,\text{arcsec}$ across.
Integrated fluxes for \textit{grizY}, \textit{J}, \textit{K}, \textit{W1} and \textit{W2} bands are indicated in the coloured circles.
The observed fluxes measured by the three surveys are indicated by the black squares with error bars.
The BICs for the model fits are given in the legend.
For J0109--5424, a galaxy+quasar template provides a significantly better fit than a pure galaxy or quasar template, and the source is discernable in the \textit{g} band imaging (top row, far left).
For J0603--3923 and J0122--4609, the best-fitting galaxy+quasar models are almost identical to the best-fitting quasar models, and there is no clear source in the \textit{g} band.
}
\label{fig:fits}
\end{figure*}

\subsection{The cool dwarf J0043--6028}

The final object in `Quasar Island' is J0043--6028, which also had archival unpublished EFOSC2/NTT spectroscopy (programme 0100.A-0346(A), PI: Reed), taken over $1800\,\text{s}$ using grating $\#16$ on the night of 2018 November 13/14.
The achieved depth of these observations was too low to conclude on the nature of this object.
Therefore, we re-observed this object using EFOSC2/NTT on the night of 2023 November 30 / December 1 (programme 112.25VZ.001, PI: Belladitta) for $7200\,\text{s}$ using grating $\# 5$.
The data were reduced using \textsc{Pypeit} \citep{pypeit:joss_pub} using the associated bias, flats and standard GD108 (observed during the night with the same grating). 

We show the extracted 1D spectrum in the final panel of Fig.~\ref{fig:all_spectra}.
The 1D spectrum is consistent with that of a M9 dwarf star \citep{Burgasser2008}, consistent with the DES brown dwarf census using DES Y3, VHS and WISE photometry \citep{CarneroRosell2019}.

\section{Discussion} \label{discussion}

\subsection{Evaluation of contrastive learning methodology}

Of the 11 objects selected by the contrastive network, 10 are high-redshift quasars and one is a cool dwarf star, giving an efficiency (true positive rate) of 91 per cent.
At least three high-redshift quasars were not selected, giving a miss rate of at least 23 per cent.
Our methodology therefore has not identified all of the quasars in the sample; it is possible that further high-redshift quasars in the DES--VHS--WISE footprint remain undiscovered.
Other machine learning methods may be able to provide a more complete selection.

We have managed to dispense with an extensive visual inspection stage in our candidate selection.
Although not excessively time-consuming for $\sim10^2$ objects, this would have been tedious with our much larger working sample of $7\,438$, obtained with the broader selection criteria we require to include lensed quasars.
Aside from requiring significant expert time, visual inspection procedures are unlikely to be entirely systematic.
The lack of this stage in our methodology therefore aids its reproducibility.

\subsection{J0109--5424 as a candidate lensed high-redshift BAL quasar} \label{SED}
\label{sec:lensed_quasar}
Whilst the Lyman-$\alpha$ line and redwards continuum of J0603--3923 is typical of high-redshift quasars, J0109--5424 presents strong absorption features and weak Lyman-$\alpha$ emission (see Fig.~\ref{fig:all_spectra}).
The strong absorption at $\sim 9100-9800\,\text{\AA}$ is consistent with \ion{Si}{IV} broad absorption line (BAL) outflow.
The shape of the BAL is strongest at high velocities (up to $\sim 18000\ \rm{km\ s}^{-1}$), which is unusual in BALs at high- and low-redshifts \citep[e.g.,][]{Weymann1991, Trump2006, Bischetti2022}.
A near-infrared spectrum confirming the presence of the associated \ion{C}{IV} BAL and/or contamination of the presumed \ion{Si}{IV} BAL by other absorption features, as well as characterising the shape of the rest-frame UV continuum up to and beyond the \ion{C}{IV} line, is necessary to conclude.
We also identify a \ion{C}{IV} absorption system at $z=4.655$, although a higher resolution spectrum would be necessary to confirm it.

Most interestingly, the spectrum of J0109--5424 presents significant flux throughout the Lyman-$\alpha$ forest and even below the Lyman limit.
This is an unambiguous signature of a foreground, potentially lensing, galaxy contaminating the photometry of the quasar, making J0109--5424 the first lensed BAL quasar candidate at $z>5.5$. Further, the \textsc{SExtractor} \textit{class\_star} probability (0: extended source, 1: point-source) for the \textit{g} band is $0.32$, suggesting that the contaminating object is likely a galaxy\footnote{
    For the \textit{rizY} bands, the \textit{class\_star} probabilities are respectively $0.82$, $0.96$, $0.98$, $0.84$, consistent with the point-source appearance of the quasar in these bands.
}.

SED modelling also suggests J0109--5424 to be lensed, unlike J0603--3923 or J0122--4609.
We fit the photometry of these three quasars to templates of galaxies, quasars, and lensed quasars, using a custom code employing the \textsc{emcee} Python package \citep{emcee}.
The galaxy models were constant-star-formation-rate models from the \textsc{BAGPIPES} package \citep{carnall18}, parametrized by stellar mass, start and end times of star formation, metallicity, redshift, and dust extinction $A_\textit{V}$.
The quasar templates were generated using \textsc{QSOGEN} \citep{Temple2021,temple22}, parametrized using a flux-free scaling factor, colour excess $E(B-V)$, and the IGM absorption prescription from \citet{Becker2013}\footnote{
    We note that \textsc{QSOGEN} uses the IGM absorption prescription of \citet{Becker2013} which results in bluer $i-z$ colours for $z>5.8$ quasars (e.g., due to less Lyman-$\alpha$ forest absorption) than observed \citep[see Appendix B of ][ for more details]{schindler23}.
    In our specific case, this makes the modelling of a combined foreground galaxy and background quasar spectrum more conservative as the inclusion of a lensing galaxy would also make the $i-z$ colour bluer.
}.
Finally, we create simple `galaxy+quasar' composite templates by adding the galaxy templates to the quasar templates, in order to mimic to first order the expected spectrum of a lensed quasar system.
The galaxy templates have six free parameters; the quasar templates two; the composite lensed quasar templates eight.
To quantify goodness of fit for models with different numbers of fitting parameters, we use the Bayesian information criterion (BIC; \citealt{schwarz78}), which penalizes models with more parameters to discourage overfitting.
The best-fitting spectra for the three quasars against the three templates are shown in Fig.~\ref{fig:fits}, along with the BICs for the each candidate and model.
For J0109--5424, we find that a galaxy+quasar model is a significantly better model than a pure quasar model by $\Delta\text{BIC}>150$.
The left panel of Fig.~\ref{fig:fits} shows that the inclusion of galaxy flux in the modelling adds sufficient flexibility to fit the flux measured in the \textit{g} band.
The foreground galaxy redshift is poorly constrained, which is unsurprising considering the galaxy is not outshone by the quasar only in the \textit{g} band. The redshift distribution is bimodal with two peaks at $z=0.4$ and $z=1.3$, in agreement with the most probable lens redshifts for $z=6$ quasars \citep{pacucci19,yue22}. We note that the maximum-likelihood foreground galaxy model underpredicts the strong emission observed at $5800\,\text{\AA} <\lambda<6300\,\text{\AA}$.
Full spectroscopic and photometric modelling of the quasar-galaxy composite spectrum is however outside of the scope of this paper and would require deeper and higher-resolution spectroscopy.
For the other two objects, the best-fitting galaxy+quasar model is essentially identical to the best-fitting quasar model: the galaxies in these models contribute very little flux across all bands.
These quasars have a higher BIC for galaxy+quasar models than pure quasar models; we therefore consider these objects to be unlensed quasars.

Another argument for the lensed nature of J0109--5424 is its extremely short proximity zone.
Using the standard definition of the ionized near-zone \citep{Eilers2021}, we measure $r_\text{NZ}= 0.41 \pm 0.12(\pm 1.15) \,\rm{Mpc}$, where the first error is due to the resolution of the GMOS spectra and the second stems from the uncertainty in the quasar redshift.
The expected proximity zone for J0109--5424 ($M_{1450} = -26.52$) is $r_\text{NZ}\simeq 4 \,\rm{Mpc}$, although considerable scatter exists in the $z\sim6$ quasar population \citep{Eilers2017,Eilers2021,Satyavolu2023}.
Assuming the best-fit model of \citet{Satyavolu2023} for the population of $6<z<6.2$ quasars (excluding 'young' quasars with short proximity zones from \citealt{Eilers2020}), the $1\sigma$ upper limit on the measured proximity zone of J0109--5424 $r_\text{NZ}<1.56\ \rm{Mpc}$ would be expected for a $M_{1450} = 23.5$ quasar. 
The discrepancy with the apparent magnitude then strongly argues for this object to be gravitationally lensed and magnified (see e.g. \citealt{Davies2020}), with an implied total magnification of $\mu \simeq 16$.
Another alternative explanation of the short proximity zone is that J0109--5424 is a young quasar.
Indeed, the size of its proximity zone is consistent with models and observations of quasars with short lifetimes $t_q \simeq 10^4\,\text{yrs}$ \citep{Eilers2020, Satyavolu2023}.
A more precise redshift (for example using FIR lines targeted with ALMA or rest-frame optical lines with \textit{JWST}) is needed together with a higher-resolution optical spectrum to improve the proximity zone measurement.

In summary, multiple arguments point to a lensed nature for J0109~{--}5424: the detection of flux below the Lyman limit; the SED modelling; the near-zone size (assuming the quasar is not young).
Space-based high-resolution imaging is necessary to definitely confirm the lensed nature of J0109--5424 by detecting multiple images.

\subsection{Implications for the lensing fraction at \texorpdfstring{$z\sim 6$}{TEXT}}

Confirming the lensed nature of J0109--5424 would double the observed lensed fraction of $z\sim 6$ quasars.
The DES--VHS--WISE detected sample we have used in this work contains $13$ $z\sim6$ quasars (see Section \ref{contrastive}), of which one is a lensed quasar candidate.
This would imply an observed lensed fraction of $7.7_{-4.7}^{+10.8}$ per cent.
These estimates assume that the VHS-WISE detection criteria do not preferentially select lensed or non-lensed quasars. A more pessimistic estimate can be obtained by using all $z\sim 6$ quasars in the DES footprint passing the $i-z$ and $z-Y$ cuts detailed in Section \ref{dataprep}. This would result in $21$ quasars, giving a lensed fraction of $4.2^{+6.1}_{-2.6}$ per cent.
Both estimates are consistent with \citet{pacucci19} who predict a $z\sim 6$ lensed fraction between $6$ and $20$ per cent, and \citet{yue22} who give a revised estimate using more recent lens parameters at $1$--$6$ per cent depending on the bright-end slope of the $z\sim 6$ quasar luminosity function and the depth of the survey considered. 
Compared to the several hundred high-redshift quasars discovered to date, the lensed fraction we obtain predicts $\sim 10$ quasars at $z\sim 6$ to be lensed, suggesting that several are yet to be discovered in archival ground-based imaging survey data.

\section{Conclusions} \label{conc}

We have developed a novel methodology based on unsupervised machine learning to find high-redshift quasars that have been missed by previous surveys, including gravitationally-lensed objects.
This new method, using relaxed colour cuts and DES imaging data, is completely automated and does not require an extensive visual inspection stage.
Our method isolated 11 objects in the DES--VHS--WISE footprint, 10 of which are high-redshift quasars, implying an efficiency of $91$ per cent.
We note however that the selection is not complete: three known high-redshift quasars are located outside of `Quasar Island'.
Of these, we report the discovery of three new $z\sim 6$ quasars.
One object, J0109--5424 at $z=6.07$, is likely strongly lensed: it is detected in the \textit{g} band, its spectrum shows significant flux in the Lyman-$\alpha$ forest and even below the Lyman limit, and the ionized near-zone is too small considering the apparent luminosity of the quasar.
Detection in the \textit{g} band had likely caused this quasar to be missed by previous searches.
Follow-up high-resolution imaging is now needed to confirm the lensed nature of J0109--5424, which would make it the second lensed $z>6$ quasar to be discovered.

The discovery of a lensed $z>6$ quasar in the DES footprint would suggest that of order 10 similar quasars have been missed in previous searches due to the photometric contamination induced by lensing galaxies.
The magnification effect induced by gravitational lensing brings ever fainter quasars into view, and this work suggests that lensed high-redshift quasars are not as uncommon as the paucity of the known population would suggest.
As such it is likely that lensed quasars can be observed deeper into the epoch of reionisation, opening up new avenues with which to probe the high-redshift Universe.

\section*{Acknowledgements}
The authors thank the anonymous referee for comments and suggestions which improved this work.
This research made use of the cross-match service provided by CDS, Strasbourg \citep{boch12}.
In addition to Python packages referenced in the text, we also acknowledge the use of \textsc{NumPy} \citep{numpy}, \textsc{Matplotlib} \citep{matplotlib}, \textsc{PyVO} \citep{pyvo}, \textsc{pandas} \citep{pandas1, pandas2}, \textsc{SciPy} \citep{scipy} and \textsc{Astropy} \citep{astropy1, astropy2, astropy3}.

XB thanks the Max Planck Institute for Astronomy, for the funding of the internship within which this work was conducted.
RAM, FW acknowledge support from the ERC Advanced Grant 740246 (Cosmic\_Gas).
RAM acknowledges support from the Swiss National Science Foundation (SNSF) through project grant 200020\_207349.
EPF is supported by the international Gemini Observatory, a program of NSF’s NOIRLab, which is managed by the Association of Universities for Research in Astronomy (AURA) under a cooperative agreement with the National Science Foundation, on behalf of the Gemini partnership of Argentina, Brazil, Canada, Chile, the Republic of Korea, and the United States of America.
RD and LF acknowledge support from the INAF GO 2022 grant ``The birth of the giants: JWST sheds light on the build-up of quasars at cosmic dawn''. RD acknowledges support by the PRIN MUR ``2022935STW''.

Based on observations (Programme ID GS-2022B-FT-208) obtained at the international Gemini Observatory, a program of NSF’s NOIRLab, which is managed by the Association of Universities for Research in Astronomy (AURA) under a cooperative agreement with the National Science Foundation on behalf of the Gemini Observatory partnership: the National Science Foundation (United States), National Research Council (Canada), Agencia Nacional de Investigación y Desarrollo (Chile), Ministerio de Ciencia, Tecnología e Innovación (Argentina), Ministério da Ci\^encia, Tecnologia, Inovaç\~oes e Comunicaç\~es (Brazil), and Korea Astronomy and Space Science Institute (Republic of Korea).

Also based on observations collected at the European Southern Observatory under ESO programmes 0100.A-0345(A), 098.A-0439(A), and 112.25VZ.001. The authors thank S. Reed and R. McMahon for proposing and observing programmes 098.A-0439(A) and 0100.A-0345(A).

\section*{Data Availability}

The DES photometric and imaging data used in this work are publicly available at \url{https://www.darkenergysurvey.org/the-des-project/data-access/}.
The contrastive network architecture and trained weights will be made available upon acceptance of the manuscript at \url{https://github.com/xbyrne/glq_mpia}.
The NTT/EFOSC2 raw data of J0122--4609 and J0043--6028 presented in this work are available in the ESO archive (Proposal IDs: 098.A-0439(A), 112.25VZ.001).
The Gemini GMOS spectra raw data are available in the Gemini Observatory Archive (Programme ID GS-2022B-FT-208).



\bibliographystyle{mnras}
\bibliography{bibliography}




\appendix

\section{\texttt{SQL} query} \label{app_sql}

Here we present the \texttt{SQL} code used to query the DES database.

\begin{verbatim}
SELECT 
    COADD_OBJECT_ID, RA, DEC,
    MAG_APER_4_G, MAG_APER_4_R, MAG_APER_4_I, 
    MAG_APER_4_Z, MAG_APER_4_Y,
    MAGERR_APER_4_G, MAGERR_APER_4_R,
    MAGERR_APER_4_I, MAGERR_APER_4_Z, 
    MAGERR_APER_4_Y
FROM DR2_MAGNITUDE WHERE
    MAG_APER_4_Z < 21.0 AND
    MAGERR_APER_4_Z < 0.1 AND
    MAG_APER_4_Y < 22.45 AND
    MAGERR_APER_4_Y < 0.1 AND
    MAG_APER_4_I - MAG_APER_4_Z > 1.0 AND
    FLAGS_G < 4 AND FLAGS_R < 4 AND FLAGS_I < 4 AND
    FLAGS_Z < 4 AND FLAGS_Y < 4 AND
    IMAFLAGS_ISO_G = 0 AND IMAFLAGS_ISO_R = 0 AND 
    IMAFLAGS_ISO_I = 0 AND IMAFLAGS_ISO_Z = 0 AND 
    IMAFLAGS_ISO_Y = 0
\end{verbatim}

\section{Contrastive Learning Training \& Architecture} \label{app_contrastive}

The hyperparameters used in training the contrastive learning model are given in Table~\ref{tab:hyperparameters}.
The architecture of the model is given in Table~\ref{tab:architecture}.
We arrived at this architecture and these hyperparameters by starting with those of \citet{sarmiento21} and modifying until most of the known high-redshift quasars appeared on a separate island.
Ablative tests showed that four convolutional layers (as in \citealt{sarmiento21}) were unnecessary, but two layers were insufficient to distinguish the high-redshift quasars from the cool dwarf stars; we therefore used three layers.
We found that doubling or halving any of the numbers of filters in the convolutional layers from the values given in Table~\ref{tab:architecture} did not improve separation; likewise the numbers of neurons in the dense layers.
The learning rate and momentum were similarly optimised until most of the known high-redshift quasars were successfully separated.

The network was trained using the cloud-based GPUs offered by Google Colaboratory\footnote{
\url{https://colab.research.google.com}
}.
Early stopping was implemented with a patience of 5 and a minimum delta of 0.01, such that training is halted when the loss function fails to decrease by more than 0.01 after 5 epochs;
this occurred after 88 epochs.
Each epoch took approximately $8\,\text{s}$.

\begin{table}
	\centering
	\caption{Training hyperparameters used for contrastive learning.}
	\label{tab:hyperparameters}
	\begin{tabular}{cc} 
		\hline
        Hyperparameter & Value\\
        \hline
        Batch size ($N$) & 128\\
		Temperature ($\tau$) & 0.1\\
        Optimizer & SGD\\
        Learning Rate & $3\times10^{-4}$\\
        Momentum & $1\times10^{-5}$\\
        Patience & 5\\
        Minimum Delta & 0.01\\
		\hline
	\end{tabular}
\end{table}

\begin{table}
	\centering
	\caption{
        CNN architecture.
        Layer names are those defined by TensorFlow \citep{tensorflow}.
        All Conv2D and Dense layers have exponential linear unit activation \citep{elu}.
        }
	\label{tab:architecture}
	\begin{tabular}{lllr}
		\hline
		Module
        &Layer name
        &\multicolumn{2}{l}{Hyperparameters}\\
		\hline
		Data Augmentation
            &RandomCrop&$24\times24$px\\
            &RandomFlip\\
            &RandomTranslation&$\pm2$px\\
            &RandomRotation&$\pm180^\circ$\\
        \hline
        Base Encoder
            &Conv2D & 128 filters & kernel 5\\
            &MaxPooling2D && pool 2\\
            &Conv2D & 256 & 3\\
            &MaxPooling2D && 2\\
            &Conv2D & 512 & 3\\
            &MaxPooling2D && 2\\
            &Reshape & 512\\
        \hline
        Projection Head
            &Dense & 256 neurons\\
            &Dense & 128\\
            &Dense & 128\\
		\hline
	\end{tabular}
\end{table}


\bsp	
\label{lastpage}
\end{document}